\documentclass{emulateapj}
\usepackage{hyperref,hyperref,amsmath,amssymb,times,graphicx,fleqn}

\begin{document}

\title{Suppression of stellar tidal disruption rates by anisotropic initial conditions}
\author{Kirill Lezhnin\altaffilmark{1}}
\email{klezhnin@yandex.ru, eugvas@lpi.ru}
\author{Eugene Vasiliev\altaffilmark{2,3}}
\affil{$^1$Moscow Institute of Physics and Technology, Institutskiy per. 9, Dolgoprudny, Moscow Region, Russia, 141700}
\affil{$^2$Lebedev Physical Institute, Leninsky prospekt 53, Moscow, Russia, 119991}
\affil{$^3$Rudolf Peierls Centre for Theoretical Physics, 1 Keble road, Oxford, UK, OX1 3NP}

\newcommand{\jlc}{j_\mathrm{lc}}
\newcommand{\Jgap}{J_\mathrm{gap}}
\newcommand{\Jcirc}{J_\mathrm{circ}}
\newcommand{\ah}{a_\mathrm{h}}
\newcommand{\rinfl}{r_\mathrm{infl}}
\newcommand{\Trel}{t_\mathrm{rel}}
\newcommand{\Trefill}{t_\mathrm{refill}}

\begin{abstract}
We compute the rates of capture of stars by supermassive black holes, using time-dependent Fokker--Planck equation with initial conditions that have a deficit of stars on low-angular-momentum orbits. One class of initial conditions has a gap in phase space created by a binary black hole, and the other has a globally tangentially-anisotropic velocity distribution. We find that for galactic nuclei that are younger than $\sim 0.1$ relaxation times, the flux of stars into the black hole is suppressed with respect to the steady-state value. This effect may substantially reduce the number of observable tidal disruption flares in galaxies with black hole masses $M_\bullet\gtrsim 10^7\,M_\odot$.
\end{abstract}

%%%%%%%%%%%%%%%%%%%%
\section{Introduction}

A star passing at a small enough distance from a supermassive black hole (SMBH) is captured or tidally disrupted, producing a detectable flare in multiple wavebands \citep{Rees1988}. Estimating the rate of such events is the topic of the loss-cone theory, first developed in 1970s in application to hypothetical intermediate-mass black holes in globular clusters \citep[e.g.][]{FrankRees1976, LightmanShapiro1977}. Later on, this theory was applied to SMBH in galactic centres \citep{SyerUlmer1999,MagorrianTremaine1999,WangMerritt2004}, typically estimating the tidal disruption rate to be in the range $10^{-4}-10^{-5}$ events per year per galaxy. Observational constraints from optical \citep{vanVelzenFarrar2014}, UV \citep{Gezari2008} or X-ray \citep{Donley2002, KhabibullinSazonov2014} surveys roughly fall into the same range \citep[see also a review by][]{Komossa2015}. Recently \citet{StoneMetzger2014} raised a concern that the observationally derived event rates are systematically lower than the theoretical estimates, by as much as one order of magnitude, although the uncertainties are hardly smaller than that on either side.

In this paper we investigate a mechanism that can significantly reduce the event rates in galaxies with long enough relaxation times and alleviate the tension between theory and observations, namely the influence of tangentially anisotropic initial conditions. The source of this anisotropy could be a gap in the low-angular-momentum region of the phase space, created by a binary SMBH \citep{MerrittWang2005}, or simply a mild bias towards circular orbits. We do not attempt to model individual galaxies, but focus instead on the comparison of steady-state rates, typically used in the literature, with the reduced ones from anisotropic but otherwise the same initial conditions, in order to derive the suppression factor as a function of galaxy parameters. We only consider spherically symmetric galaxies, therefore obtaining a lower boundary on the possible event rates.

The paper is organized as follows. In Section~\ref{sec:diffeqn} we write down the Fokker--Planck equation that describes the diffusion of stars in angular momentum, and derive its time-dependent analytical solution. Then in Section~\ref{sec:initcond} we present the modifications of initial conditions that have tangential anisotropy. In Section~\ref{sec:results} we obtain the estimates of capture rates for our choice of initial conditions, and compute the suppression factor with respect to the steady-state capture rates. We discuss the implications of our results in Section~\ref{sec:discussion}.

%%%%%%%%%%%%%%%%%%%%
\section{Analytical solution of the diffusion equation}  \label{sec:diffeqn}

Two-body relaxation that changes energies $E$ and angular momenta $J$ of stars can be described in terms of orbit-averaged Fokker--Planck equation \citep[e.g.][Chapter 5]{MerrittBook}. It is commonly assumed that the changes in angular momentum are much more important for computing the capture rates \citep{FrankRees1976,LightmanShapiro1977}, and for this reason the diffusion in energy is usually neglected; numerical solution of the more general two-dimensional Fokker--Planck equation \citep{CohnKulsrud1978,Merritt2015} confirms the validity of one-dimensional approximation for processes that occur on a timescale much shorter than the relaxation time $\Trel$. 
At a fixed energy $E$, the distribution function (DF) of stars $f(j)$ as a function of normalized angular momentum $j\equiv J/\Jcirc(E)$, where $J_{\rm circ}$ is the angular momentum of a circular orbit with the same energy, satisfies the diffusion equation in a cylindrical geometry:

\begin{align}  \label{eq:diffusion}
\frac{\partial f(E,j,t)}{\partial t} = 
  \frac{\mathcal{D}}{4j}\frac{\partial }{\partial j}
  \left(j \frac{\partial f}{\partial j}\right).
\end{align}
Here $\mathcal{D}(E) \propto \Trel^{-1}$ is the orbit-averaged diffusion coefficient \citep[e.g.][equation~18]{Merritt2013}, which is assumed to be independent of $j$, allowing an analytical solution to this equation \citep{MilosMerritt2003}. 
The boundary condition at $j=1$ is of the Neumann type: $\partial f/\partial j=0$ (zero-flux condition). The presence of the black hole creates a capture boundary at $j=\jlc$, the angular momentum at which a star would be tidally disrupted at periapsis.
As discussed in \citet{LightmanShapiro1977}, there are two limiting cases for the behaviour of $f(j)$ near $\jlc$, depending on the ratio $q\equiv \mathcal{D} T/\jlc^2$ between the mean-square change in $j$ due to relaxation over one orbital period $T$ and the size of the loss cone. 
In the general case, one may use a Robin-type boundary condition (a linear combination of the function and its derivative), which naturally interpolates between the regimes of empty ($q\ll1$) and full ($q\gg1$) loss cone:

\begin{align}  \label{eq:bdryinner}
f(\jlc)=\left. \frac{\alpha\jlc}2
\frac{\partial f(j)}{\partial j}\right|_{j=\jlc},\quad 
\alpha(q)\approx (q^2+q^4)^{1/4}.
\end{align} 

In the steady state, the flux of stars towards the capture boundary $\mathcal{F}\equiv (\mathcal{D}/4) j\, \partial f(j)/\partial j$ is nearly independent of $j$ at low $j$, and the solution of Equation~\ref{eq:diffusion} has a logarithmic profile: $f(j) \propto \log j + \mathrm{const}$. Extrapolating this functional form to all $j$, one obtains the classical quasi-steady-state solution \citep[e.g.][equation~51, with $\mathcal R\equiv j^2$]{VasilievMerritt2013}. \citet{CohnKulsrud1978} defined an effective capture boundary $j_0 \equiv \jlc\exp(-\alpha/2)$, at which the inward extrapolation of the logarithmic solution reaches zero; this gives the same boundary condition at the true capture boundary as Equation~\ref{eq:bdryinner}, except for their slightly different expression for $\alpha$. 

The general time-dependent solution of equation (\ref{eq:diffusion}) subject to the specified boundary conditions can be expressed in a series form \citep[equations~24--26]{MilosMerritt2003}:
\begin{align}  \label{eq:series_solution}
f(j,t) &= \sum_{m=1}^\infty C_m\,A(\beta_m,j)\,\exp(-\mathcal{D}\beta_m^2t/4) \;,\\
A(\beta,j) &\equiv J_0(\beta j) Y_1(\beta) - Y_0(\beta j) J_1(\beta) \;,\\
C_m &\equiv   \frac{(\pi^2/2)\, \beta_m^2 J_0^2(\beta_m\jlc)} {J_0^2(\beta_m\jlc)-J_1^2(\beta_m)}
  \int_{\jlc}^1\!\! j'A(\beta_m,j')\,f(j'\!,0)\,dj',\nonumber
\end{align}
where $J_i$ and $Y_i$ are Bessel functions of the first and the second kind, $A$ are the basis function, $\beta_m$ are the roots of a certain equation that satisfy the boundary conditions for each basis function, and $C_m$ are the expansion coefficients computed from the initial conditions $f(j,0)$.
The above study derived their expressions for the limiting case of an empty loss cone, i.e.\ the boundary condition $f(\jlc)=0$. Taking the same inward extrapolation of the logarithmic profile as in \citet{CohnKulsrud1978}, one can use their expressions with a modified capture boundary $j_0$ for the general case of a non-empty loss cone \citep[this method was adopted for the time-dependent solution in][]{VasilievMerritt2013}; however, the validity of this approach depends on the assumption of the solution being close to the steady-state profile. Instead, in this work we generalize the analytical solution to the arbitrary boundary conditions given by Equation~\ref{eq:bdryinner}. Namely, the coefficients $\beta_m$ that enter the expressions for the series expansion are the roots of the following equation:
\begin{align}
&\left[ J_0 (\beta_m j_{\rm lc}) + \alpha \beta_m j_{\rm lc} J_1 (\beta_m j_{\rm lc}) / 2 \right] Y_1(\beta_m) \;- 
\nonumber \\
&\left[ Y_0 (\beta_m j_{\rm lc}) + \alpha \beta_m j_{\rm lc} Y_1 (\beta_m j_{\rm lc}) / 2 \right] J_1(\beta_m)=0.
\end{align}
This modification is necessary to obtain a rigorous solution for non-trivial initial conditions, such as those considered in the next section, although it gives only a minor correction to the more approximate treatment in the case of a nearly-logarithmic form of the solution (i.e.\ late enough into the evolution).

The series solution with a finite number of terms $m_\mathrm{max}$ breaks down at small times ($t\lesssim 10\mathcal{D}^{-1}m_\mathrm{max}^{-2}$), so we compute the solution numerically with a finite-difference scheme in this case.

%%%%%%%%%%%%%%%%%%%%
\section{Tangentially anisotropic initial conditions}  \label{sec:initcond}

The stationary solution with a logarithmic $j$-dependence of DF is attained at times greater than an appreciable fraction of the relaxation time. On the other hand, this time may well exceed the age of the Universe, especially in low-density galactic nuclei. Therefore, the choice of initial conditions becomes important for determining the present-day capture rates. We consider two classes of initial DFs with a deficit of stars at low angular momenta, compared to the isotropic population.

The first possible reason for such a deficit is the ejection of stars by a binary SMBH that may have existed in a galaxy previously. The slingshot mechanism is responsible for the flattening of the density profile and the formation of cores \citep{MilosMerritt2001,Merritt2006}, which have been detected observationally \citep[e.g.][]{DulloGraham2012,Rusli2013}. More importantly, it creates a gap in the phase space, ejecting stars with angular momenta less than some critical value, which corresponds to the periapsis radius comparable to the radius of a hard binary, $\ah\equiv q/[4(1+q)^2]\,\rinfl$, where $q\equiv m_2/m_1\le 1$ is the mass ratio of the binary, and $\rinfl$ is the SMBH radius of influence. In this work we adopt the definition of $\rinfl$ as the radius containing the mass of stars equal to $2(m_1+m_2)$ before the slingshot process has started.
%%%%%%%%%%%%%%
\begin{figure}
\includegraphics{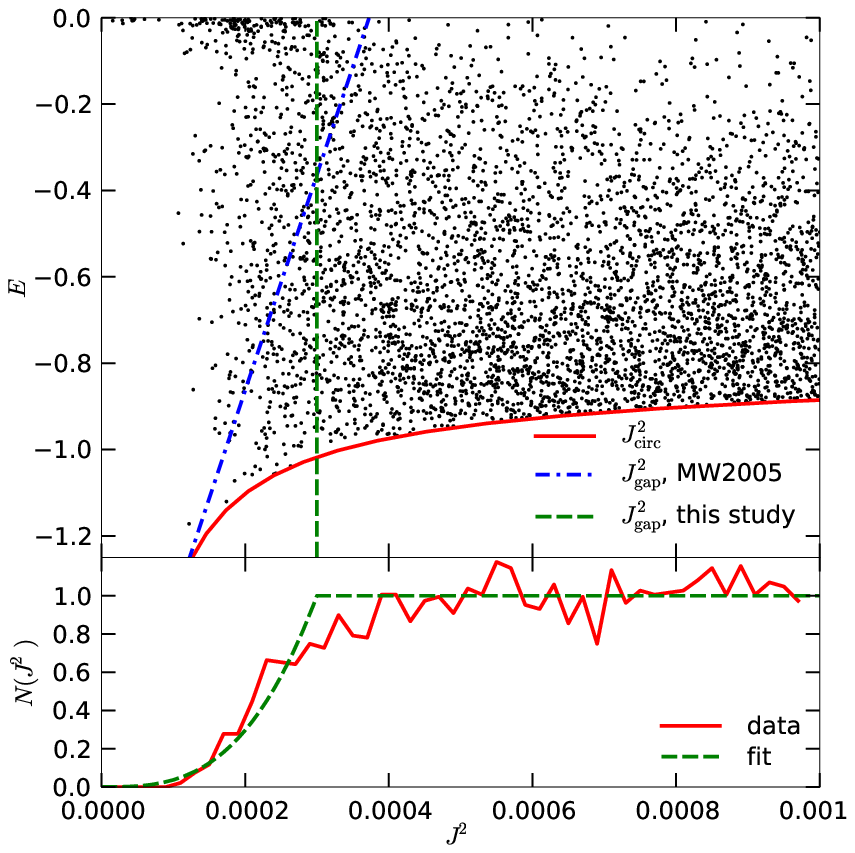}
\caption{Monte Carlo simulations of a binary SMBH in a galaxy with $\gamma=1$ Dehnen density profile. The binary mass is $10^{-2}$ of the total stellar mass, the mass ratio $q=1$, and the binary started on a nearly-circular orbit at separation 0.2, roughly equal to the radius of influence. \protect\\
\textit{Top panel:} Phase space (squared angular momentum vs.\ energy) after the binary has cleared the low angular momentum region. Dashed green and dot-dashed blue lines show the definition of the gap region in this study and in \citet{MerrittWang2005}; solid red line marks the angular momentum of a circular orbit. \protect\\
\textit{Bottom panel:} approximation of the initial distribution function (Equation~\ref{eq:df_init_gap}) used in this work.
}  \label{fig:phasespacegap}
\end{figure}
%%%%%%%%%%%%

\citet{MerrittWang2005} considered a time-dependent solution for the one-dimensional Fokker--Planck equation, using the expressions of \citet{MilosMerritt2003}, i.e.\ an empty loss cone boundary condition, and taking the initial distribution in angular momentum as a Heaviside step function: $f(E,J) \propto \Theta(J-\Jgap(E))$. They defined the gap width as $\Jgap(E)\equiv K\ah \sqrt{2\left[E-\Phi(K\ah)\right]}$, with a dimensionless constant $K\simeq 1$.
We have used a modified version of the Monte Carlo code \textsc{Raga} \citep{Vasiliev2015,VasilievAM2015} to determine the distribution of stars in angular momentum in a galaxy with a binary SMBH; our results are better fit by an energy-independent gap width, $\Jgap'\equiv \sqrt{K'\,G(m_1+m_2) \ah}$, with $K'\simeq 3$, and a more gradual drop towards smaller $J$ (Figure~\ref{fig:phasespacegap}):
\begin{align}  \label{eq:df_init_gap}
f(E,J,0) = f(E) \cdot \min \left(1, \left( \frac{J}{\Jgap'} \right)^6\right).
\end{align}

Another possible choice of initial conditions involves a DF that has a tangential anisotropy at all $J$, not just a gap at small $J$. The simplest possibility is to consider DF in a factorized form: $f(E, J) = (1-\beta) [J/\Jcirc(E)]^{-2\beta}\,\tilde f(E)$, where $\tilde f(E)$ is the counterpart of the usual isotropic DF and is computed from a given density profile with the method of \citet{Cuddeford1991}. Such DF corresponds to a constant velocity anisotropy coefficient $\beta$ \citep[equation~4.61]{BinneyTremaine}; the case of weak tangential anisotropy $\beta=-1/2$ is consistent with some observationally-based models of galactic centers \citep[e.g.][figure~2]{Thomas2014}, and results in a very simple expression for DF:
\begin{align}
\tilde f(E)=\frac{\Jcirc(E)}{3\pi^2}\:\;
\frac{d^2}{d\Phi^2}\left[\frac{\rho(\Phi)}{r(\Phi)}\right].
\end{align}
%The mass per unit energy is related to DF by $dM = \tilde f(E) g(E) dE$, where 
%\begin{align}
%g(E) &= (1-\beta) \Jcirc^{2\beta}(E)\;\; 8\pi^{5/2}\frac{\Gamma(1-\beta)}{\Gamma(3/2-\beta)}\, \\
%&\times \int_0^{r^{-1}(E)} dr\, r^{2-2\beta}\,[2(E-\Phi(r))]^{1/2-\beta}  \nonumber
%\end{align}
%(the latter formula is valid for any $\beta$).
%It turns out that $g(E)$ depends on $\beta$ very weakly (changes by no more than a few per cent), and 
%$\tilde f(E)$ for anisotropic models is also quite close to the isotropic DF, except in the region close to the black hole if the inner density slope $\gamma$ is close to $\beta+1/2$ (in which case the DF drops sharply at high $|E|$). 
%We note that galaxies with the shallowest slope of surface density profile cannot be modelled by an isotropic DF in the presence of BH, but are compatible with $\beta=-1/2$.

%%%%%%%%%%%%%%%%%%%%
\section{Results}  \label{sec:results}
We have considered a set of \citet{Dehnen1993} $\gamma$-models with a black hole mass $M_\bullet=10^{-3}$ of the total mass in stars. 
The energy-dependent part of DF $f(E)$ or $\tilde f(E)$ and the diffusion coefficient $\mathcal{D}(E)$ are computed numerically from the given density profile, using the Eddington inversion formula or its Cuddeford's generalization. 
We explore several values of the power-law index $\gamma$ of the central density profile, and for each value of $\gamma$ we chose to consider a one-parameter family of models by scaling $M_\bullet$ and $\rinfl$ simultaneously, according to the following relation \citep{MerrittSK2009}:
\begin{align}  \label{eq:Msigma}
\rinfl = r_0\,[ M_\bullet/ 10^8\,M_\odot]^{0.56} .
\end{align}
As our default normalization, we set $r_0=30$~pc, but we also consider values of $r_0=20$ and 45~pc. The ratio of the influence radius to the scale radius of Dehnen profile is \{0.091, 0.047, 0.016\} for $\gamma=\{0.5,1,1.5\}$.

It is natural to express our results in dimensionless units: the flux normalized to the steady-state capture rate, and the time measured in units of relaxation time \citep[equation~7.106]{BinneyTremaine} at $\rinfl$. For $10^6\le M_\bullet/M_\odot\le 10^8$ and $0.5\le\gamma\le1.5$, these scale approximately as
\begin{align}  \label{eq:scaling_flux}
\lg\left[\mathcal{F}_\mathrm{st}/\left(M_\odot \mathrm{yr}^{-1}\right)\right] &\approx -4.6 - 1.5\,\lg(r_0/30\;\mbox{pc})  \\
&+\: 0.2(1-\gamma)\, \lg(M_\bullet/ 10^8\,M_\odot), \nonumber 
\end{align}
\begin{align}  \label{eq:scaling_time}
\lg\left[\Trel/\mbox{yr}\right] &\approx 13 + 0.4(1-\gamma) + 1.5\,\lg(r_0/30\;\mbox{pc}) \\
&+\:1.28\,\lg (M_\bullet/ 10^8\,M_\odot). \nonumber
\end{align}

%%%%%%%%%%%%%%
\begin{figure}
\includegraphics{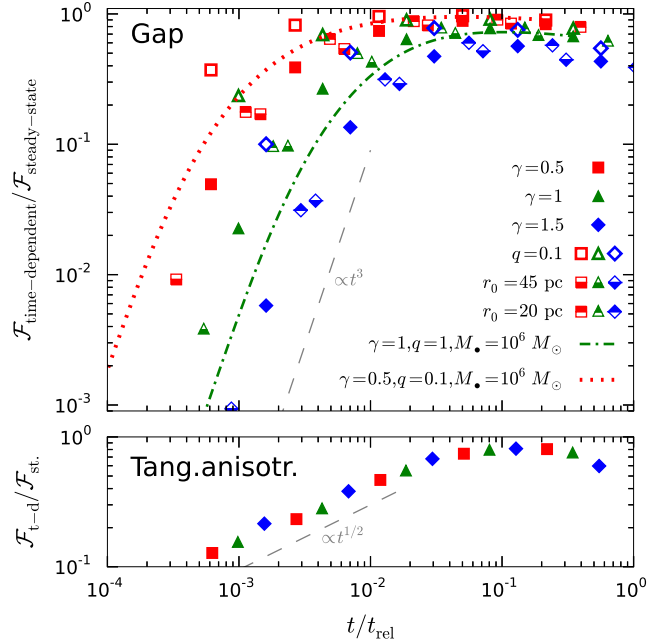}
\caption{
Suppression factor -- the ratio of time-dependent to steady-state capture rate,  as a function of time normalized to the relaxation time at the radius of influence. Individual points correspond to models with different density profiles and $M_\bullet=10^6, 10^{6.5}, \dots, 10^8\,M_\odot$; the abscissae correspond to the Hubble time ($10^{10}$~yr) measured in units of relaxation time, thus the most massive black holes are at the left side of the plot. \protect\\
\textit{Top panel}: families of models with an initial gap in angular momentum distribution due to the action of a pre-existing binary SMBH; different symbols encode $\gamma$; open symbols are for models with smaller gap width (binary mass ratio $q=1/10$) and filled -- to equal-mass binaries; half-filled symbols are for $q=1$ and different normalizations of $\rinfl-M_\bullet$ relation. Dotted and dot-dashed curves show examples of time-dependent flux for particular choices of parameters. 
\protect\\
\textit{Bottom panel}: tangentially anisotropic ($\beta=-1/2$) models for different $\gamma$.
} \label{fig:capturerate}
\end{figure}
%%%%%%%%%%%%

The steady-state flux is comparable to $M_\bullet/\Trel$.
Figure~\ref{fig:capturerate}, top panel, shows the ratio of time-dependent to steady-state capture rate, as a function of time, for two representative models with a gap. We also plot the same quantity measured at the time $10^{10}$ years, for various choices of $\gamma$, $M_\bullet$ and $q$; in this manner, the abscissa corresponds to the black hole mass according to our scaling (\ref{eq:scaling_time}). 

The time time required to establish the steady-state profile at a given energy is $\sim(\Jgap/\Jcirc)^2\,\Trel$ \citep[e.g.][equation~34]{Merritt2013}; as the maximum of the total flux arrives from energies corresponding to $\rinfl$, the time to refill the gap is roughly $\Trefill \sim (\ah/\rinfl)\, \Trel(\rinfl)$, or
\begin{align}  \label{eq:Trefill}
\Trefill \simeq 10^{13}\mbox{ yr}\times \frac{q}{4(1+q)^2} \left(\!\frac{M_\bullet}{10^8\,M_\odot}\!\right)^{\!1.28}\!
\left(\frac{r_0}{30\,\mathrm{pc}}\right)^{\!1.5} \!.
\end{align}

For $t\lesssim \Trefill$ the flux is reduced compared to the stationary value, which is commonly used in calculations of tidal disruption rates. The maximum value of capture rate reached at $t\sim \Trefill$ is somewhat lower than the steady-state value, due to the fact that the $J$-averaged DF is also depleted at high binding energies (where $\Jgap\gtrsim \Jcirc$) with respect to the value used in the steady-state calculation. The flux reaches 1/2 of its maximum value at $t_{1/2}\simeq 0.1\,\Trefill$. Moreover, at $t\gtrsim \Trefill$ it starts to decline in the absense of diffusion in energy. \citet{Merritt2015} has performed numerical integration of two-dimensional ($E,J$) Fokker--Planck equation restricted to the region inside $\rinfl$, also using initial conditions with a gap at $J<\Jgap$, and found a qualitatively similar behaviour if the diffusion in energy was artificially switched off. On the other hand, taking it into account modifies the solution at $t\gtrsim 0.1\Trel$ so that it tends to a steady-state profile. Therefore we may trust our calculations roughly up to a time when the flux reaches its maximum. 

We also explore the effect of changing the normalization in $\rinfl-M_\bullet$ relation (\ref{eq:Msigma}). This, of course, modifies both the time-dependent flux and the relaxation time at $\rinfl$ (\ref{eq:scaling_flux},\ref{eq:scaling_time}), but the normalized values still stay on the same curve for each $\gamma$ and $q$.
Finally, the second class of models with globally tangentially anisotropic initial conditions (Figure~\ref{fig:capturerate}, bottom panel) produce a milder decline of the capture rate at $t\lesssim 0.1\,\Trel$.

%%%%%%%%%%%%%%%%%%%
\section{Discussion and conclusions}  \label{sec:discussion}

We have considered the question whether the capture rate of stars by SMBH can be substantially lowered with respect to the steady-state value by a suitable modification of initial conditions. We used an analytical time-dependent solution of the Fokker--Planck equation to compute the capture rate for the given initial conditions as a function of time. Two classes of initial conditions were analyzed: a gap in the low-angular-momentum region of phase space, created by a binary SMBH, and a separable distribution function with a mild tangential velocity anisotropy. 

The results of our study can be summarized as follows. If the relaxation time in the galaxy centre is short enough ($\lesssim 10^{11}$~yr), any difference between initial conditions is erased quite rapidly, and the capture rate approaches the steady-state value. On the other hand, for $\Trel\gtrsim 10^{12}$~yr anisotropic initial conditions may substantially reduce the capture rate (in the intermediate range of $\Trel$, their effect is moderate). In our series of models with a phase-space gap, for SMBH masses $M_\bullet \gtrsim 10^7\,M_\odot$ the suppression factor (the ratio of time-dependent to steady-state capture rates) drops quite rapidly, plunging below $10^{-1}$ for $M_\bullet \gtrsim 10^8\,M_\odot$. Note that for even heavier black holes, visible flares constitute a small fraction of the captured stars \citep[e.g.][figure~15]{MacleodGR2012}. In models with mild tangential anisotropy the suppression factor is not so extreme, but nevertheless can reduce the flux by a factor of a few for $M_\bullet\gtrsim 10^8\,M_\odot$. 

We have extended the work of \citet{MerrittWang2005} into the range of smaller SMBH masses ($\le 10^8\,M_\odot$), since they are more numerous in the Universe and are expected to dominate the overall tidal disruption rates. Moreover, smaller $M_\bullet$ residing in more compact galactic nuclei are not well described by the empty-loss-cone regime adopted in that paper. In this study we have derived the solution for the general case; we checked that assuming an empty-loss-cone boundary condition does not substantially change the results for $M_\bullet \gtrsim 10^7\,M_\odot$, but overestimates the flux by a factor of a few for the least massive SMBHs. We used a somewhat different initial distribution function inside the gap than the above paper, but checked that adopting their initial conditions changes the results only marginally. However, our estimates of the time required for the capture rate to reach 1/2 of its steady-state value, $t_{1/2}\simeq 0.1\Trefill$, the latter given by Equation~\ref{eq:Trefill}, are about an order of magnitude longer than shown in figure~3 or equation~14 of \citet{MerrittWang2005} for the same galaxy parameters\footnote{We thank D.Merritt for providing us the original data from the 2005 paper.}. We believe that this discrepancy might be due to a calibration error in that paper, as our expressions for $\Trefill$ agree with equation~7 in \citet{MerrittSzell2006} and equation~36 in \citet{Merritt2013}. 

We deliberately have made a number of simplifying assumptions that drive our capture rates towards lower values. First, we assumed a spherical geometry, while it is known that non-spherical torques can result in a higher efficiency of loss-cone repopulation \citep{MagorrianTremaine1999,MerrittPoon2004,HolleySigurdsson2006}. Naturally, the difference between spherical and non-spherical geometry starts to manifest itself above the same threshold value of $M_\bullet$ as the influence of initial conditions \citep[figure~4 in][]{Vasiliev2014}, again underlining the distinction between relaxed and non-relaxed galactic nuclei. Note that by relaxed we here mean the systems that had enough time to establish a nearly steady-state logarithmic profile in angular momentum distribution; this does not mean they were able to relax in energy space as well and develop a \citet{BahcallWolf1976} cusp. Second, we neglected non-classical phenomena that may increase the relaxation rate \citep[e.g.][]{Alexander2012}, such as mass segregation \citep{FreitagAK2006}, massive perturbers \citep{PeretsHA2007}, or resonant relaxation \citep[e.g.][Chapter 5.6]{MerrittBook}; the latter, however, is typically not very effective in boosting the capture rates \citep{HopmanAlexander2006}. A number of other refinements have been shown by \citet{StoneMetzger2014} to have little impact on the final values. Binary SMBH that have not yet coalesced also suppress the capture rates \citep{ChenLM2008}, although they may demonstrate brief episodes of increased encounter rates at early stage of evolution \citep{IvanovPS2005,ChenSML2011,WeggBode2011}. Thus our findings can be regarded as robust lower limits on the capture rates for single SMBH. 

The effect investigated in this paper is unlikely to substantially reduce the rate of tidal disruption events for black hole masses smaller than $\sim 10^7\,M_\odot$. Although the volumetric rate of tidal disruption events is dominated by galaxies with smallest $M_\bullet$ \citep[e.g.\ figure~8 in][]{StoneMetzger2014}, the rate of \textit{observable} events (figure~10 in that paper) may be significantly affected by the suggested mechanism if either (1) the SMBH mass function is suppressed at low-mass end, or (2) flare emission mechanisms are inefficient for low-mass SMBH. These factors are still the main sources of uncertainty in the estimates of the observable event rate \citep{StoneMetzger2014}; however, it is notable that the distribution of actually observed flares peaks around $M_\bullet=10^7\,M_\odot$ (figure~11 in that paper). Thus the effect considered here may potentially be important in comparing the theoretical predictions of tidal disruption rates with observations.

KL is supported by Dynasty Foundation through the Scholarship Program for Students in Theoretical Physics.
EV acknowledges support from NASA (grant No.\ NNX13AG92G).
This work was partially supported by Russian Foundation for Basic Research (grant No.\ 15-02-03063).
We thank David Merritt and Nick Stone for valuable comments, and the referee for constructive feedback. The software for computing the capture rates is available at \url{http://td.lpi.ru/~eugvas/losscone}.

\label{lastpage}

\end{document}